\documentclass[english]{article}
\usepackage[T1]{fontenc}
\usepackage[latin9]{inputenc}
\usepackage{float}
\usepackage{amstext}
\usepackage{stackrel}
\usepackage{graphicx}
\usepackage{esint}

\makeatletter

\pdfoutput=1
\usepackage[T1]{fontenc}
\usepackage[latin9]{inputenc}
\usepackage{color}
\usepackage{float}
\usepackage{graphicx}
\usepackage{esint}
\usepackage{enumerate}

\makeatletter

\usepackage{amsfonts,amssymb}
\usepackage{latexsym}
\usepackage{epsfig}
\usepackage{cite}
\usepackage{graphicx}
\usepackage[colorlinks,linkcolor=blue]{hyperref}
\newcommand{\be}{\begin{equation}}
\newcommand{\ee}{\end{equation}}

\usepackage{babel}

\makeatother

\usepackage{babel}
\begin{document}
{}~ \hfill\vbox{\hbox{ }}\break
\vskip 3.0cm
\centerline{\Large \bf  Resolving naked singularities in $\alpha^{\prime}$-corrected string theory}

\vspace*{10.0ex}
\centerline{\large Shuxuan Ying}
\vspace*{7.0ex}
\vspace*{4.0ex}
\centerline{\large \it Department of Physics, Chongqing University}
\vspace*{1.0ex}
\centerline{\large \it Chongqing, 401331, China} \vspace*{1.0ex}
\vspace*{4.0ex}

\centerline{ysxuan@cqu.edu.cn}
\vspace*{10.0ex}
\centerline{\bf Abstract} \bigskip \smallskip

Low energy effective action of bosonic string theory possesses a kind of   singular static solution which can be interpreted as a naked singularity. Based on the Hohm-Zwiebach action, the naked singularities  could be smoothed out by introducing the complete $\alpha^{\prime}$ corrections of string theory. In this paper, we present two sets of non-singular solutions, which are also regular everywhere in the Einstein frame. In the perturbative region $\alpha^{\prime}\to0$, the solutions reduce to the  perturbative results.   Our result provides extra evidence for weak cosmic censorship conjecture (WCCC) from a viewpoint of string theory.

\vfill
\eject
\baselineskip=16pt
\vspace*{10.0ex}
\tableofcontents

\section{Introduction}

T-duality of string theory has been studied for a long time. It presents
a physical equivalence of moving strings in two compactified backgrounds
with radii $R$ and $\alpha^{\prime}/R$, where $\alpha^{\prime}$
denotes square of the string length. In low energy effective theory
of closed strings, for certain configurations, background's compactification
is not necessary and T-duality can be replaced by a close duality,
the scale-factor duality \cite{Tseytlin:1991wr,Veneziano:1991ek,Meissner:1991zj, Sen:1991zi,Sen:1991cn,Tseytlin:1991xk}.
The scale-factor duality was first discovered in the equations of
motion (EOM) of the closed string's low energy effective action. It
shows that the EOM in FLRW cosmological background are invariant under
a transformation: $a\left(t\right)\longleftrightarrow1/a\left(t\right)$.
Einstein's gravity does not possess this property since a closed string
dilaton $\phi$ plays a central role in this duality. In addition,
the scale-factor duality brings us a remarkable pre-big bang cosmology
\cite{Veneziano:2000pz,Gasperini:2002bn,Gasperini:2007vw,Gasperini:1992em}.
It implies that our universe was not created from an initial big bang
singularity, but from a stage of evolution before the big bang. Pre-
and post- big bang scenarios are disconnected due to the big bang
singularity, which could be removed by higher loop corrections \cite{Gasperini:1992em,Gasperini:2003pb,Gasperini:2004ss, Gasperini:book}.

In 1997, Meissner found that when massless closed string fields only
depended on time, zeroth and first order $\alpha^{\prime}$ corrections
of the low energy effective action could be simplified and rewritten
in $O\left(d,d\right)$ invariant forms \cite{Meissner:1996sa}. After
that, Sen proved that all orders in $\alpha^{\prime}$ are $O(d,d)$
invariant\cite{Sen:1991cn,Sen:1991zi}. By assuming that the standard
$O\left(d,d\right)$ form could be maintained, Hohm and Zwiebach \cite{Hohm:2015doa,Hohm:2019ccp,Hohm:2019jgu}
demonstrated that all order $\alpha^{\prime}$ corrections of the
low energy effective action could be dramatically simplified in the
FLRW cosmological background, namely the Hohm-Zwiebach action. Since
the Hohm-Zwiebach action is a polynomial function of the standard
$O\left(d,d\right)$ matrix, the EOM only contain first two order
derivative terms. This result made the whole theory stable (ghost
free) and solvable. The higher-order derivative terms and $O\left(d,d\right)$
breaking terms could be eliminated by a suitable field redefinition.
Moreover, it turns out that non-perturbative dS vacua could be allowed
in bosonic string theory. Based on these works, we and collaborators
generalized the Hohm-Zwiebach action to a braneworld ansatz and provided
a possibility of an existence of non-perturbative AdS vacua \cite{Wang:2019mwi}.
Then, we figured out non-singular string cosmological solutions of
the Hohm-Zwiebach action, which showed that the big bang singularity
could be smoothed out by $\alpha^{\prime}$ corrections \cite{Wang:2019kez}.
The systematic method to construct the solutions of the Hohm-Zwiebach
action were shown in our following work \cite{Wang:2019dcj}. Moreover,
since the EOM of the Hohm-Zwiebach action only contains the first
two derivative terms of the metric, which meets the requirement of
Lovelock gravity, we presented an equivalence between the Hohm-Zwiebach
action and Lovelock gravity in cosmological background by appropriate
field redefinitions \cite{Wang:2020eln}. On the other hand, there
is still some work trying to introduce $O\left(d,d\right)$ covariant
matter sources into the Hohm-Zwiebach action \cite{Bernardo:2019bkz}
and double field theory \cite{Wu:2013sha,Lescano:2021nju}.

Since the Hohm-Zwiebach action has been generalized from the cosmological
ansatz to the braneworld ansatz \cite{Wang:2019mwi}, it is possible
to discuss the naked singularities in this model. Naturally, these
new kinds of solutions will include non-perturbative effects of string
theory, and then help us understand the weak cosmic censorship conjecture
(WCCC) \cite{Penrose:1969pc} from the viewpoint of string theory.
In this paper, we start with the tree-level closed string's low energy
effective action. The braneworld solutions of this action possess
the naked singularity. The previous discussions of naked singularities
in bosonic string theory were presented in ref. \cite{Kar:1998rv}.
Based on our previous work \cite{Wang:2019dcj}, we provide two classes
of non-singular and non-perturbative solutions, which match perturbative
solutions to an arbitrary order in $\alpha^{\prime}$. It is worth
noting that the term non-perturbative here means the solutions are
valid everywhere in all regions of $|y|$  and $\alpha^{\prime}$,
in contrast to that the perturbative solutions only exist in $|y|\to\infty$
(or equivalently $\alpha^{\prime}\to0$). Our results show that non-perturbative
$\alpha^{\prime}$ corrections can smooth out the naked singularities
of spacetime. In addition, we wish to stress that our work could be
used to the braneworld scenarios in the near future, which was born
from discussions and explorations of the theory of extra dimensions.
The first step it to study how to define the junction functions, tensor
perturbations and graviton localization in the Hohm-Zwiebach action.
Good reviews for braneworld models can be found in refs. \cite{Dzhunushaliev:2009va,Liu:2017gcn}.

The reminder of this paper is outlined as follows. In section 2, we
present naked singular solutions of the tree-level closed string's
low energy effective action. In section 3, based on the Hohm-Zwiebach
action, we provide two classes of non-singular solutions. Finally,
we present some inspirations and conclusions in section 4.

\section{Naked singularities in tree-level effective action}

We start with the low energy gravi-dilaton effective action of string
theory. For simplicity, we only consider the gravi-dilaton system
without any perfect fluid sources, and set anti-symmetric Kalb-Ramond
field $b_{ij}=0$. The action is given by

\begin{equation}
S=\frac{1}{2\kappa^{2}}\int d^{d+1}x\sqrt{-g}e^{-2\phi}\left[R+4\left(\partial_{\mu}\phi\right)^{2}\right],\label{eq:original action}
\end{equation}

\noindent where $D=d+1$ is the spacetime dimension, $\phi$ is the
closed string dilaton and $g_{\mu\nu}$ is the string metric which
relates to the Einstein metric through $g_{\mu\nu}^{E}=\mathrm{exp}\left(-\frac{4}{D-2}\phi\right)g_{\mu\nu}$.
By varying the action (\ref{eq:original action}), the equations of
motion (EOM) are obtained as

\noindent
\begin{eqnarray}
R_{\mu\nu}+2\nabla_{\mu}\nabla_{\nu}\phi & = & 0,\nonumber \\
\nabla^{2}\phi-2\left(\partial_{\mu}\phi\right)^{2} & = & 0.\label{eq:EoM without potential}
\end{eqnarray}

\noindent And then, we introduce the $O\left(d,d\right)$ invariant
dilaton field $\Phi$ which is defined by the closed string dilaton
$\phi$ and string metric $g_{\mu\nu}$:

\begin{equation}
\Phi=2\phi-\ln\sqrt{-g}.
\end{equation}

\noindent The braneworld ansatz is given by

\begin{equation}
ds^{2}=\bar{a}^{2}\left(y\right)\eta_{\mu\nu}dx^{\mu}dx^{\nu}+dy^{2},\label{eq:brane world ansatz}
\end{equation}

\noindent where $\bar{a}\left(y\right)$ denotes a wrap factor of
braneworld theory and $\eta_{\mu\nu}=\mathrm{diag}\left(-1,1\text{,1,}1,\ldots\right)$
is the metric of Minkowski spacetime. The EOM of (\ref{eq:EoM without potential})
become

\begin{eqnarray}
2\Phi^{\prime\prime}-\Phi^{\prime2}-d\bar{H}^{2} & = & 0,\nonumber \\
-d\bar{H}^{2}+\Phi^{\prime\prime} & = & 0,\nonumber \\
\bar{H}^{\prime}-\Phi^{\prime}\bar{H} & = & 0,\label{eq:EoM sim}
\end{eqnarray}

\noindent where we used a notation $f^{\prime}\left(y\right)\equiv\partial_{y}f\left(y\right)$
for an arbitrary function $f\left(y\right)$, and the definition of
the Hubble-like parameter is given by $\bar{H}\left(y\right)\equiv\frac{\bar{a}^{\prime}\left(y\right)}{\bar{a}\left(y\right)}$.
It is noteworthy that $\bar{H}\left(y\right)$ can be used to show
the features of background curvature of (\ref{eq:brane world ansatz}).
It is easy to check that the EOM (\ref{eq:EoM sim}) are invariant
under the scale-factor duality:

\begin{equation}
\bar{a}\left(y\right)\longleftrightarrow\bar{a}\left(y\right)^{-1},\qquad\bar{H}\left(y\right)\longleftrightarrow-\bar{H}\left(y\right),\qquad\Phi\left(y\right)\longleftrightarrow\Phi\left(y\right),\label{eq:scale factor duality}
\end{equation}

\noindent and possess a $Z_{2}$ symmetry: $y\rightarrow-y$. The
self-dual solutions are

\begin{equation}
\bar{a}_{\pm}\left(y\right)=\left|\frac{y}{y_{0}}\right|^{{ \pm 1}/\sqrt{d}},\qquad\bar{H}_{\pm}\left(y\right)=\pm\frac{1}{\sqrt{d}y},\qquad\Phi\left(y\right)=-\ln\left|\frac{y}{y_{0}}\right|,\label{eq:leading solution}
\end{equation}

\noindent where $y_{0}$ is an integral constant. This result also
can be transformed to the solutions in the spherical coordinates \cite{Kar:1998rv}.
The curvature properties of these solutions can be presented by Hubble-like
parameters $\bar{H}_{\pm}\left(y\right)$ in Fig. (\ref{fig:perturbative}):

\begin{figure}[H]
\noindent \begin{centering}
\includegraphics[scale=0.4]{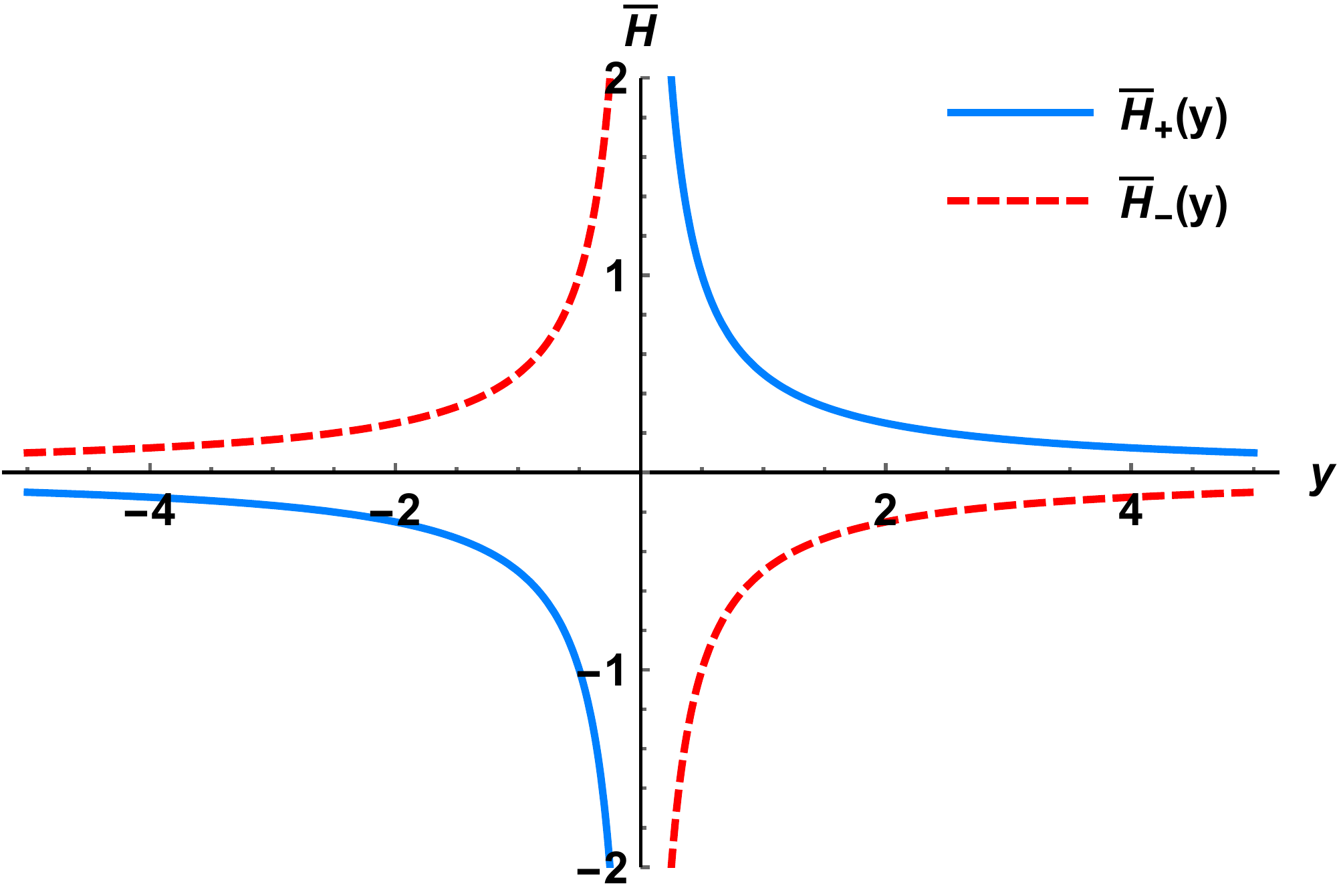}
\par\end{centering}
\centering{}\caption{\label{fig:perturbative} Hubble-like parameters $\bar{H}$ vs. coordinate
of  dimension $y$ of four solutions (we set $d=4$ in this plot).}
\end{figure}

Since the EOM (\ref{eq:EoM sim}) are invariant under the scale-factor
duality (\ref{eq:scale factor duality}), the action (\ref{eq:original action})
can manifest this invariance. To present this invariance, let us define
a new quantity

\begin{equation}
g_{\mu\nu}=\left(\begin{array}{cc}
1 & 0\\
0 & \bar{G}_{\mu\nu}\left(y\right)
\end{array}\right),\label{eq:set up 1}
\end{equation}

\noindent based on the following ansatz:

\begin{equation}
ds^{2}=\bar{G}_{\mu\nu}\left(y\right)dx^{\mu}dx^{\nu}+dy^{2}.
\end{equation}

\noindent Using this new notation, the action (\ref{eq:original action})
can be rewritten as

\begin{equation}
\bar{I}_{0}=\int dxe^{-\Phi}\left[\Phi^{\prime2}+\frac{1}{8}\mathrm{Tr}\left(\bar{\mathcal{S}}^{\prime2}\right)\right].\label{eq:0th action}
\end{equation}

\noindent where the $O\left(d,d\right)$ standard metric $\bar{\mathcal{S}}$
is defined by a symmetric $2d\times2d$ matrix $\bar{M}$
\[
\bar{M}=\left(\begin{array}{cc}
\bar{G}^{-1} & 0\\
0 & \bar{G}
\end{array}\right),
\]
and

\begin{equation}
\bar{\mathcal{S}}=\eta\bar{M}=\left(\begin{array}{cc}
0 & \bar{G}\\
\bar{G}^{-1} & 0
\end{array}\right),\label{eq:ODD matrix}
\end{equation}

\noindent in which $\eta$ is the invariant metric of the $O\left(d,d\right)$
group

\begin{equation}
\eta=\left(\begin{array}{cc}
0 & I\\
I & 0
\end{array}\right).
\end{equation}

\noindent Furthermore, the action (\ref{eq:0th action}) is invariant
under the $O\left(d,d\right)$ transformation:

\begin{equation}
\Phi\rightarrow\Phi,\qquad\bar{\mathcal{S}}\rightarrow\tilde{\bar{\mathcal{S}}}=\Omega^{T}\bar{\mathcal{S}}\Omega,
\end{equation}

\noindent where $\Omega$ is a constant matrix that satisfies

\begin{equation}
\Omega^{T}\eta\Omega=\eta.
\end{equation}

\noindent It is worth noting that the $O\left(d,d\right)$ invariant
transformation

\begin{equation}
\bar{\mathcal{S}}=\left(\begin{array}{cc}
0 & \bar{G}\\
\bar{G}^{-1} & 0
\end{array}\right)\longleftrightarrow\tilde{\bar{\mathcal{S}}}=\left(\begin{array}{cc}
0 & \bar{G}^{-1}\\
{\normalcolor \bar{G}} & 0
\end{array}\right),
\end{equation}

\noindent is the same as the scale-factor duality $\bar{a}\left(y\right)\longleftrightarrow\bar{a}\left(y\right)^{-1}$
in eq. (\ref{eq:scale factor duality}) when we utilize the braneworld
ansatz (\ref{eq:brane world ansatz}), say $\bar{G}_{\mu\nu}\left(y\right)=\bar{a}^{2}\left(y\right)\eta_{\mu\nu}$.

\section{Non-singular solutions via $\alpha^{\prime}$ corrections}

In this section, let us demonstrate how to obtain the regular solutions
by introducing the $\alpha^{\prime}$ corrections to the gravi-dilaton
system in the string frame (\ref{eq:original action}). To include
higher order $\alpha^{\prime}$ corrections, the action (\ref{eq:original action})
can be rewritten as

\begin{equation}
S=\int d^{d+1}x\sqrt{-g}e^{-2\phi}\left[R+4\left(\partial_{\mu}\phi\right)^{2}+\frac{\alpha^{\prime}}{4}R_{\mu\nu\rho\sigma}R^{\mu\nu\rho\sigma}+\ldots\right].\label{eq:corrected action}
\end{equation}

\noindent In ref. \cite{Zwiebach:1985uq}, Zwiebach showed that the
first order $\alpha^{\prime}$ correction could be rewritten as a
well-known Gauss-Bonnet term by a suitable field redefinition:

\begin{equation}
S=\int d^{d+1}x\sqrt{-g}e^{-2\phi}\left[R+4\left(\partial_{\mu}\phi\right)^{2}+\frac{\alpha^{\prime}}{4}\left(R_{\mu\nu\rho\sigma}R^{\mu\nu\rho\sigma}-4R_{\mu\nu}R^{\mu\nu}+R^{2}\right)+\ldots\right].
\end{equation}

\noindent Using the $O\left(d,d\right)$ invariant matrix (\ref{eq:ODD matrix})
and applying a further field redefinition \cite{Meissner:1996sa},
this action can be simplified

\begin{equation}
\bar{I}=-\int dxe^{-\Phi}\left[-\Phi^{\prime2}-\frac{1}{8}\mathrm{Tr}\bar{\mathcal{S}}^{\prime2}-\frac{\alpha^{\prime}}{4}\left(\frac{1}{16}\mathrm{Tr}\bar{\mathcal{S}}^{\prime4}-\frac{1}{32}\left(\mathrm{Tr}\bar{\mathcal{S}}^{\prime2}\right)^{2}\right)+\ldots\right].
\end{equation}

\noindent Based on Sen's proof \cite{Sen:1991cn,Sen:1991zi} and the
action above, Hohm and Zwiebach conjectured that all orders in $\alpha^{\prime}$
corrections could be rewritten in terms of the $O\left(d,d\right)$
invariant matrix $\bar{\mathcal{S}}$ in eq. (\ref{eq:ODD matrix}).
Considering all possible combinations of $\bar{\mathcal{S}}$ and
its trace and derivatives for higher-order terms, and using a set
of field redefinitions, the action can be dramatically simplified
as follows \cite{Hohm:2019jgu,Wang:2019mwi}:

\begin{equation}
\bar{I}_{HZ}=-\int dxe^{-\Phi}\left(-\Phi^{\prime2}+\stackrel[k=1]{\infty}{\sum}\left(\alpha^{\prime}\right)^{k-1}\bar{c}_{k}\mathrm{Tr}\left(\bar{\mathcal{S}}^{\prime2k}\right)+\mathrm{multi-traces}\right),\label{eq: Odd with alpha x}
\end{equation}

\noindent in which we define a new set of coefficients $\bar{c}_{k}$'s.
The new coefficients have a relationship with $c_{k}$'s in Hohm-Zweibach
action as below:

\begin{equation}
\bar{c}_{1}=c_{1},\qquad\bar{c}_{2k+1}=c_{2k+1},\qquad\bar{c}_{2k}=-c_{2k},\qquad k=1,2,\ldots,
\end{equation}

\noindent where $c_{1}=-\frac{1}{8}$, $c_{2}=\frac{1}{64}$ for the
bosonic string and $c_{k\geq3}$ are undetermined constants which
could be fixed by string theory in future works. After substituting
the braneworld ansatz (\ref{eq:brane world ansatz}) into the action
(\ref{eq: Odd with alpha x}), we have

\begin{equation}
\bar{I}_{HZ}=-\int dxe^{-\Phi}\left(-\Phi^{\prime2}-d\sum_{k=1}^{\infty}\left(-\alpha^{\prime}\right)^{k-1}2^{2k+1}\bar{c}_{k}\bar{H}^{2k}\right).\label{eq:HZ action}
\end{equation}

\noindent The complete derivation for the action (\ref{eq: Odd with alpha x})
can be found in the appendix of ref. \cite{Wang:2019mwi}. The EOM
of (\ref{eq:HZ action}) can be obtained directly, which are

\begin{eqnarray}
\Phi^{\prime\prime}+\frac{1}{2}\bar{H}\bar{f}\left(\bar{H}\right) & = & 0,\nonumber \\
\frac{d}{dy}\left(e^{-\Phi}\bar{f}\left(\bar{H}\right)\right) & = & 0,\nonumber \\
\Phi^{\prime2}+\bar{g}\left(\bar{H}\right) & = & 0,\label{eq:EOM}
\end{eqnarray}

\noindent where

\begin{eqnarray}
\bar{f}\left(\bar{H}\right) & = & d\stackrel[k=1]{\infty}{\sum}\left(-\alpha^{\prime}\right)^{k-1}2^{2\left(k+1\right)}k\bar{c}_{k}\bar{H}^{2k-1},\nonumber \\
\bar{g}\left(\bar{H}\right) & = & d\stackrel[k=1]{\infty}{\sum}\left(-\alpha^{\prime}\right)^{k-1}2^{2k+1}\left(2k-1\right)\bar{c}_{k}\bar{H}^{2k}.\label{eq:EOM fh gh}
\end{eqnarray}

\noindent Obviously, the EOM (\ref{eq:EOM}) reduce to (\ref{eq:EoM sim})
when $\alpha^{\prime}$ corrections disappear. Moreover, we have the
following conditions:

\begin{equation}
\bar{g}^{\prime}\left(\bar{H}\right)=\bar{H}\bar{f}^{\prime}\left(\bar{H}\right),\quad{\rm and}\quad\bar{g}\left(\bar{H}\right)=\bar{H}\bar{f}\left(\bar{H}\right)-\int_{0}^{\bar{H}}\bar{f}\left(y\right)dy,\label{eq:gf relation}
\end{equation}
It is straightforward to check that the EOM (\ref{eq:EOM}) are invariant
under the $O\left(d,d\right)$ transformations:

\begin{equation}
\bar{H}\left(y\right)\rightarrow-\bar{H}\left(y\right),\qquad\Phi\left(y\right)\rightarrow\Phi\left(y\right),\qquad\bar{f}\left(y\right)\rightarrow-\bar{f}\left(y\right),\qquad\bar{g}\left(x\right)\rightarrow\bar{g}\left(y\right).
\end{equation}

Now, let us try to figure out non-singular and non-perturbative solutions
of the EOM (\ref{eq:EOM}). Since the non-perturbative solutions must
match the perturbative solutions to an arbitrary order in $\alpha^{\prime}$
as $|y|\to\infty$ (or equivalently $\alpha^{\prime}\to0$), we will
first calculate the perturbative solutions to an arbitrary order in
the next subsection. Then, we will provide a systematic method to
construct the non-singular and non-perturbative solutions which match
the perturbative solutions.

\subsection{Perturbative solution}

We now discuss the perturbative solutions of the EOM (\ref{eq:EOM})
based on ref. \cite{Hohm:2019jgu}. For convenient, we introduce a
new variable $\Omega$ as
\begin{equation}
\Omega\equiv e^{-\Phi},
\end{equation}
where $\Omega^{\prime}=-\Phi^{\prime}\Omega$ and $\Omega^{\prime\prime}=\left(-\Phi^{\prime\prime}+\Phi^{\prime2}\right)\Omega$.
And the EOM (\ref{eq:EOM}) become
\begin{eqnarray}
\Omega^{\prime\prime}-\bar{h}\left(\bar{H}\right)\Omega & = & 0,\nonumber \\
\frac{d}{dy}\left(\Omega\bar{f}\left(\bar{H}\right)\right) & = & 0,\nonumber \\
\Omega^{\prime2}+\bar{g}\left(\bar{H}\right)\Omega^{2} & = & 0,\label{eq:reEOM}
\end{eqnarray}
where we define a new function

\begin{equation}
\bar{h}\left(\bar{H}\right)\equiv\frac{1}{2}\bar{H}\bar{f}\left(\bar{H}\right)-\bar{g}\left(\bar{H}\right)=\alpha^{\prime}2^{5}d\bar{c}_{2}H^{4}+\ldots,
\end{equation}

\noindent with

\begin{eqnarray}
\bar{f}\left(\bar{H}\right) & = & 2^{4}d\bar{c}_{1}\bar{H}-\alpha^{\prime}4\cdot2^{5}d\bar{c}_{2}\bar{H}^{3}+\ldots,\nonumber \\
\bar{g}\left(\bar{H}\right) & = & 2^{3}d\bar{c}_{1}\bar{H}^{2}-\alpha^{\prime}3\cdot2^{5}d\bar{c}_{2}\bar{H}^{4}+\ldots.
\end{eqnarray}

\noindent It is easy to see that $\bar{h}\left(\bar{H}\right)=0$
at zeroth order of $\alpha^{\prime}$. Now, we assume the perturbative
solutions of the EOM (\ref{eq:reEOM}) take forms:

\begin{eqnarray}
\Omega\left(y\right) & = & \Omega_{0}\left(y\right)+\alpha^{\prime}\Omega_{1}\left(y\right)+\alpha^{\prime2}\Omega_{2}\left(y\right)+\ldots,\nonumber \\
\bar{H}\left(y\right) & = & \bar{H}_{0}\left(y\right)+\alpha^{\prime}\bar{H}_{1}\left(y\right)+\alpha^{\prime2}\bar{H}_{2}\left(y\right)+\ldots.
\end{eqnarray}

\noindent where we denote $\Omega_{i}$ and $\bar{H}_{i}$ as the
$i$-th order of the perturbative solutions. Based on these perturbative
solutions, functions $\bar{h}\left(\bar{H}\right)$, $\bar{f}\left(\bar{H}\right)$
and $\bar{g}\left(\bar{H}\right)$ become

\begin{eqnarray}
\bar{h}\left(\bar{H}\right) & = & \alpha^{\prime}2^{5}d\bar{c}_{2}H_{0}^{4}+\ldots,\nonumber \\
\bar{f}\left(\bar{H}\right) & = & 2^{4}d\bar{c}_{1}\bar{H}_{0}+\alpha^{\prime}2^{4}d\left(\bar{c}_{1}\bar{H}_{1}-8\bar{c}_{2}\bar{H}_{0}^{3}\right)+\ldots,\nonumber \\
\bar{g}\left(\bar{H}\right) & = & 2^{3}d\bar{c}_{1}\bar{H}_{0}^{2}+\alpha^{\prime}2^{4}d\left(\bar{c}_{1}\bar{H}_{0}\bar{H}_{1}-6\bar{c}_{2}\bar{H}_{0}^{4}\right)+\ldots.
\end{eqnarray}

\noindent Then, we can substitute these perturbative forms back into
the EOM (\ref{eq:reEOM}), and solve the differential equations at
each order of $\alpha^{\prime}$ to get $\Omega_{i}$ and $\bar{H}_{i}$.
For example, the EOM (\ref{eq:reEOM}) at zeroth order in $\alpha^{\prime}$
give :

\begin{eqnarray}
\Omega_{0}^{\prime\prime} & = & 0,\nonumber \\
\frac{d}{dy}\left(2^{4}d\bar{c}_{1}\bar{H}_{0}\Omega_{0}\right) & = & 0,\nonumber \\
\Omega_{0}^{\prime2}+2^{3}d\bar{c}_{1}\bar{H}_{0}^{2}\Omega_{0}^{2} & = & 0.\label{eq:0th EOM}
\end{eqnarray}

\noindent Based on the first equation of (\ref{eq:0th EOM}), the
solution is
\begin{equation}
\Omega_{0}\left(y\right)=\gamma\left(y-\sigma_{0}\right),\label{eq:0th omega}
\end{equation}
where $\gamma$ and $\sigma_{0}$ are integral constants. Then, considering
the second and third equations of the EOM (\ref{eq:0th EOM}), the
solution of $\bar{H}_{0}$ is
\begin{equation}
\bar{H}_{0\pm}\left(y\right)=\mp\frac{\mathrm{sgn}\left(y\right)}{\sqrt{d}}\frac{1}{y-\sigma_{0}}.\label{eq:0th H}
\end{equation}
Next, the EOM (\ref{eq:reEOM}) at first order of $\alpha^{\prime}$
give

\begin{eqnarray}
\alpha^{\prime}\Omega_{1}^{\prime\prime}-\alpha^{\prime}2^{5}d\bar{c}_{2}H_{0\pm}^{4}\Omega_{0} & = & 0,\nonumber \\
\frac{d}{dy}\left(\alpha^{\prime}2^{4}d\bar{c}_{1}\bar{H}_{0\pm}\Omega_{1}+\alpha^{\prime}2^{4}d\left(\bar{c}_{1}\bar{H}_{1\pm}-8\bar{c}_{2}\bar{H}_{0\pm}^{3}\right)\Omega_{0}\right) & = & 0,\nonumber \\
2\alpha^{\prime}\Omega_{0}^{\prime}\Omega_{1}^{\prime}+\alpha^{\prime}2^{4}d\bar{c}_{1}\bar{H}_{0\pm}^{2}\Omega_{0}\Omega_{1}+\alpha^{\prime}2^{4}d\left(\bar{c}_{1}\bar{H}_{0\pm}\bar{H}_{1\pm}-6\bar{c}_{2}\bar{H}_{0\pm}^{4}\right)\Omega_{0}^{2} & = & 0.\label{eq:1st EOM}
\end{eqnarray}

\noindent Based on the first equation of (\ref{eq:1st EOM}) and a
zeroth-order solution (\ref{eq:0th omega}), we will get
\begin{equation}
\Omega_{1}\left(y\right)=\frac{2^{4}\bar{c}_{2}\gamma}{d}\frac{1}{y-\sigma_{0}}.
\end{equation}
Combining the second and third equations of the EOM (\ref{eq:1st EOM}),
and using solutions at zeroth-order (\ref{eq:0th omega}) and (\ref{eq:0th H}),
we have
\begin{equation}
\bar{H}_{1\pm}\left(y\right)=\mp\frac{80\mathrm{sgn}\left(y\right)c_{2}}{d^{3/2}}\frac{1}{\left(y-\sigma_{0}\right)^{3}}.
\end{equation}

\noindent Therefore, we obtain the first two orders in perturbative
solutions of the EOM (\ref{eq:reEOM})

\begin{eqnarray}
\bar{H}_{\pm}\left(y\right) & = & \mp\frac{\mathrm{sgn}\left(y\right)}{\sqrt{d}}\frac{1}{y-\sigma_{0}}\mp\alpha^{\prime}\frac{80\mathrm{sgn}\left(y\right)c_{2}}{d^{3/2}}\frac{1}{\left(y-\sigma_{0}\right)^{3}}+\ldots,\nonumber \\
\Omega\left(y\right) & = & \gamma\left(y-\sigma_{0}\right)+\alpha^{\prime}\frac{2^{4}c_{2}\gamma}{d}\frac{1}{y-\sigma_{0}}+\ldots.
\end{eqnarray}

\noindent Based on these methods, in the perturbative regime $|y|\to\infty$
(or equivalently $\alpha^{\prime}\to0$), the EOM can be solved iteratively
to arbitrary order in $\frac{\sqrt{\alpha^{\prime}}}{y}$ by using
eq. (\ref{eq:EOM fh gh}). For simpliciy, we set the integral constant
$\sigma_{0}=0$, and define $y_{0}\equiv\frac{\sqrt{\alpha^{\prime}}}{\sqrt{2d}}$,
the solutions with positive sign become:

\begin{eqnarray}
\bar{H}\left(y\right) & = & \frac{\sqrt{2}}{\sqrt{\alpha^{\prime}}}\left[\frac{y_{0}}{y}-160\bar{c}_{2}\frac{y_{0}^{3}}{y^{3}}+\frac{256\left(770\bar{c}_{2}^{2}+19\bar{c}_{3}\right)}{3}\frac{y_{0}^{5}}{y^{5}}\right.\nonumber \\
 &  & \left.-\frac{2048\left(88232\bar{c}_{2}^{3}+4644\bar{c}_{3}\bar{c}_{2}+41\bar{c}_{4}\right)}{5}\frac{y_{0}^{7}}{y^{7}}+\mathcal{O}\left(\frac{y_{0}^{9}}{y^{9}}\right)\right],\nonumber \\
\Phi\left(y\right) & = & -\frac{1}{2}\log\left(\beta^{2}\frac{y^{2}}{y_{0}^{2}}\right)-32\bar{c}_{2}\frac{y_{0}^{2}}{y^{2}}+\frac{256\left(44\bar{c}_{2}^{2}+\bar{c}_{3}\right)}{3}\frac{y_{0}^{4}}{y^{4}}\nonumber \\
 &  & -\frac{2048\left(6976\bar{c}_{2}^{3}+352\bar{c}_{3}\bar{c}_{2}+3\bar{c}_{4}\right)}{15}\frac{y_{0}^{6}}{y^{6}}+\mathcal{O}\left(\frac{y_{0}^{8}}{y^{8}}\right),\label{eq:perturbative solution}
\end{eqnarray}
where the second solution can be given by $\Phi=-\log\Omega$ and
\begin{eqnarray}
\bar{f}\left(\bar{H}\left(y\right)\right) & = & -2d\bar{H}-128\bar{c}_{2}d\alpha^{\prime}\bar{H}^{3}+768\bar{c}_{3}d\alpha^{\prime2}\bar{H}^{5}-4096\bar{c}_{4}d\alpha^{\prime3}\bar{H}^{7}+\mathcal{O}\left(\alpha^{\prime4}\bar{H}^{9}\right),\nonumber \\
 & = & \frac{\sqrt{d}}{y_{0}}\left[-\frac{2y_{0}}{y}+64\bar{c}_{2}\frac{y_{0}^{3}}{y^{3}}-\frac{512\left(50\bar{c}_{2}^{2}+\bar{c}_{3}\right)}{3}\frac{y_{0}^{5}}{y^{5}}\right.\nonumber \\
 &  & \left.+\frac{4096\left(2632\bar{c}_{2}^{3}+124\bar{c}_{3}\bar{c}_{2}+\bar{c}_{4}\right)}{5}\frac{y_{0}^{7}}{y^{7}}+\mathcal{O}\left(\frac{y_{0}^{9}}{y^{9}}\right)\right],\nonumber \\
\bar{g}\left(\bar{H}\left(y\right)\right) & = & -dH^{2}-96\bar{c}_{2}d\alpha^{\prime}\bar{H}^{4}+640\bar{c}_{3}d\alpha^{\prime2}\bar{H}^{6}-3584\bar{c}_{4}d\alpha^{\prime3}\bar{H}^{8}+\mathcal{O}\left(\alpha^{\prime4}\bar{H}^{10}\right),\nonumber \\
 & = & \frac{1}{y_{0}^{2}}\left[-\frac{y_{0}^{2}}{y^{2}}+128\bar{c}_{2}\frac{y_{0}^{4}}{y^{4}}-\frac{2048\left(50\bar{c}_{2}^{2}+\bar{c}_{3}\right)}{3}\frac{y_{0}^{6}}{y^{6}}\right.\nonumber \\
 &  & \left.+\frac{8192\left(24448\bar{c}_{2}^{3}+1136\bar{c}_{3}\bar{c}_{2}+9\bar{c}_{4}\right)}{15}\frac{y_{0}^{8}}{y^{8}}+\mathcal{O}\left(\frac{y_{0}^{10}}{y^{10}}\right)\right],
\end{eqnarray}

\noindent where $\beta$ is an integration constant. Moreover, the
dual solutions $-\bar{H}\left(y\right)$, $\Phi\left(y\right),$$-\bar{f}\left(y\right)$
and $\bar{g}\left(y\right)$ are also included in eq. (\ref{eq:perturbative solution}).
It is obvious that the leading term of solutions (\ref{eq:perturbative solution})
covers the result of (\ref{eq:leading solution}). And solutions possess
the curvature singularity located at $\left|y\right|\rightarrow0$.

\subsection{Non-perturbative solutions}

Before starting the discussion on the non-perturbative solutions in
this section, let us focus on a property of the EOM (\ref{eq:EOM})
at first. Based on the EOM (\ref{eq:EOM}), we find that all functions
$\bar{g}\left(y\right)$, $\bar{f}\left(y\right)$ and $\bar{H}\left(y\right)$
can be uniquely determined by $\Phi\left(y\right)$:
\begin{eqnarray}
\bar{g}\left(\bar{H}\left(y\right)\right) & = & -\Phi^{\prime2},\nonumber \\
\bar{f}\left(\bar{H}\left(y\right)\right) & = & \frac{2\sqrt{2}d}{\sqrt{\alpha^{\prime}}}\beta e^{\Phi},\nonumber \\
\bar{H}\left(y\right) & = & \frac{\sqrt{\alpha^{\prime}/2}}{\beta d}\frac{\Phi^{\prime\prime}}{e^{\Phi}},\label{eq:EOM in dilaton}
\end{eqnarray}
where $\beta$ is an integration constant. These equations (\ref{eq:EOM in dilaton})
imply that we only need to choose a proper $\Phi\left(y\right)$ to
get the non-singular solutions. In other words, it suffices to find
a regular function of $\Phi\left(y\right)$ which can be expanded
as (\ref{eq:perturbative solution}) in the perturbative regime $|y|\to\infty$
(or equivalently $\alpha^{\prime}\to0$). Then the solutions of $\bar{g}\left(y\right)$,
$\bar{f}\left(y\right)$ and $\bar{H}\left(y\right)$ will be automatically
satisfied based on eq. (\ref{eq:EOM in dilaton}). In the following
discussions, we will present two possible constructions for $\Phi\left(y\right)$,
and give their corresponding non-singular solutions $\bar{g}\left(y\right)$,
$\bar{f}\left(y\right)$ and $\bar{H}\left(y\right)$.

\subsection*{Solution A}

The first solution is given by

\begin{equation}
\Phi\left(y\right)=\frac{1}{2}\log\left(\sum_{k=1}^{\infty}\frac{\lambda_{k}}{1+\gamma^{2k}}\right),\qquad\gamma\equiv\frac{y}{y_{0}}=\frac{\sqrt{2d}}{\sqrt{\alpha^{\prime}}}y.\label{eq:ansatz A}
\end{equation}

\noindent From eq. (\ref{eq:EOM in dilaton}), we get

\begin{eqnarray}
\bar{H}\left(y\right) & = & \frac{-\left(\sum\limits _{k=1}^{\infty}\frac{2k\lambda_{k}\gamma^{2k-1}}{\left(\gamma^{2k}+1\right)^{2}}\right)^{2}+\left(\sum\limits _{k=1}^{\infty}\frac{\lambda_{k}}{\gamma^{2k}+1}\right)\sum\limits _{k=1}^{\infty}\left(\frac{8k^{2}\lambda_{k}\gamma^{4k-2}}{\left(\gamma^{2k}+1\right)^{3}}-\frac{2k(2k-1)\lambda_{k}\gamma^{2k-2}}{\left(\gamma^{2k}+1\right)^{2}}\right)}{\sqrt{2}\sqrt{\alpha^{\prime}}\beta\left(\sum\limits _{k=1}^{\infty}\frac{\lambda_{k}}{\gamma^{2k}+1}\right)^{5/2}},\nonumber \\
\bar{f}\left(\bar{H}\left(y\right)\right) & = & -\frac{2\sqrt{2}\beta d}{\sqrt{\alpha^{\prime}}}\sqrt{\sum_{k=1}^{\infty}\frac{\lambda_{k}}{\gamma^{2k}+1}},\nonumber \\
\bar{g}\left(\bar{H}\left(y\right)\right) & = & -\frac{2d\left(\sum\limits _{k=1}^{\infty}\frac{k\lambda_{k}\gamma^{2k-1}}{\left(\gamma^{2k}+1\right)^{2}}\right)^{2}}{\alpha^{\prime}\left(\sum\limits _{k=1}^{\infty}\frac{\lambda_{k}}{\gamma^{2k}+1}\right)^{2}}.\label{eq:general H}
\end{eqnarray}

\begin{figure}[H]
\noindent \begin{centering}
\includegraphics[scale=0.45]{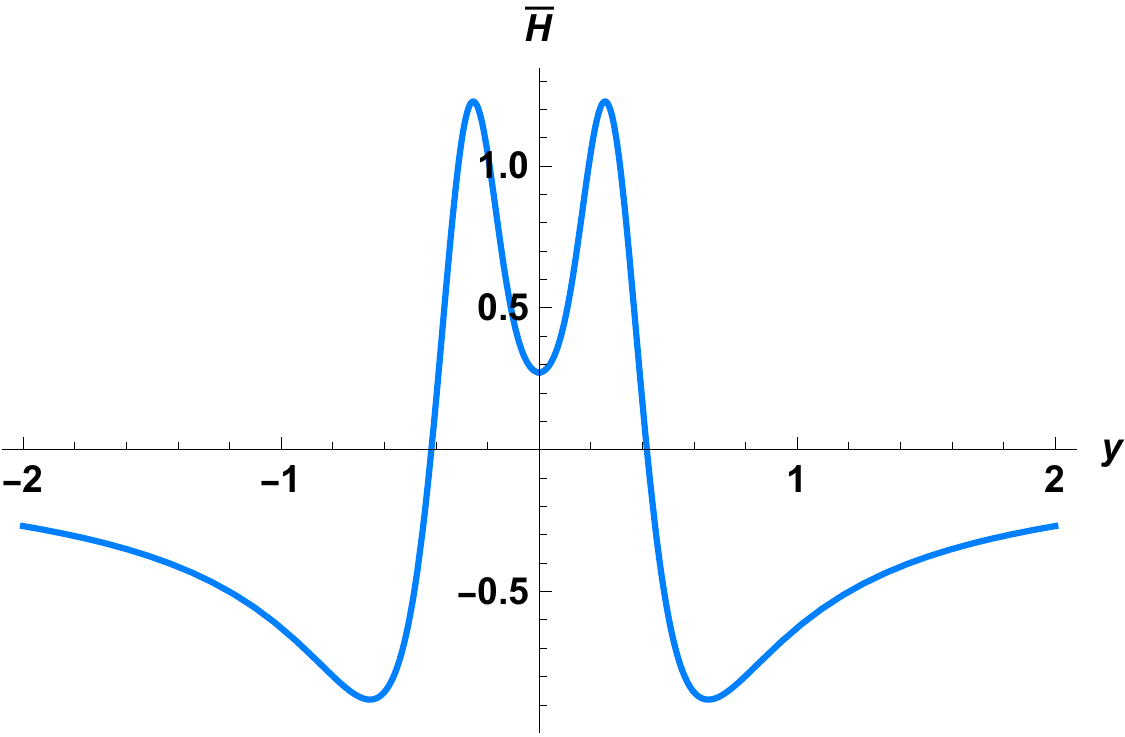}$\qquad$\includegraphics[scale=0.45]{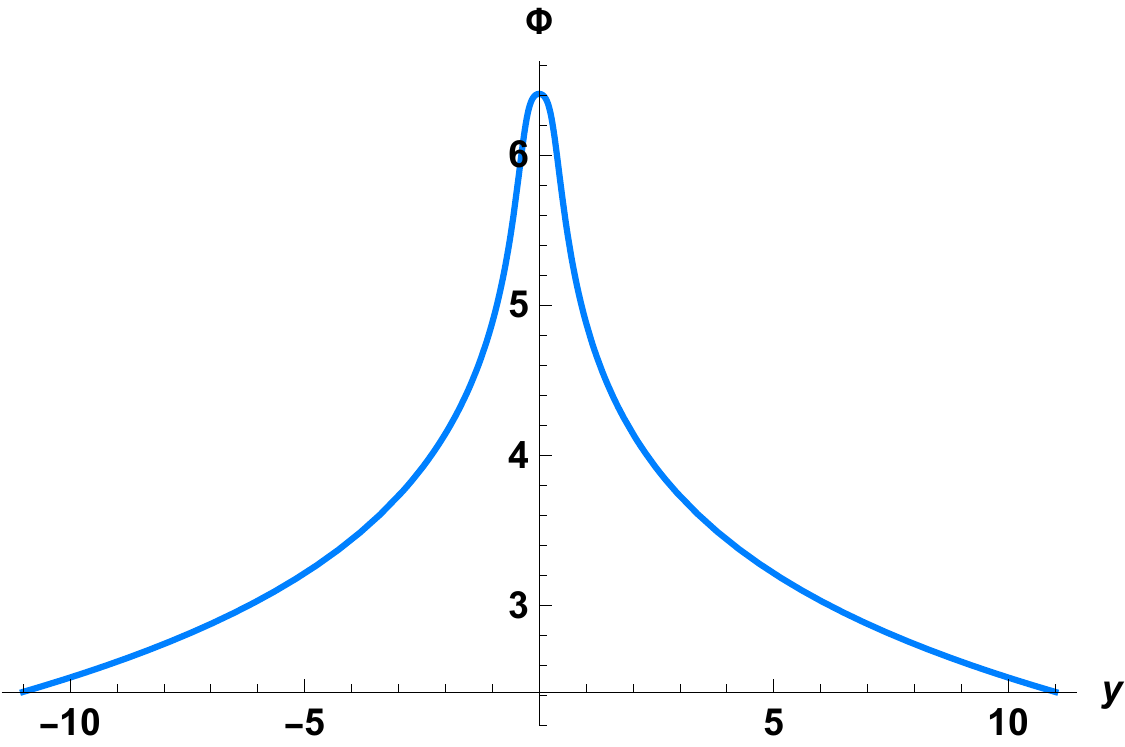}
\par\end{centering}
\caption{The left panel shows the non-perturbative Hubble-like parameters $\bar{H}\left(y\right)$
vs. coordinate of dimension $y$. The right panel presents the $O\left(d,d\right)$
dilaton $\Phi\left(y\right)$ along the dimension $y$. We set parameters
$d=4$ and $\alpha^{\prime}=1$.\label{fig:solution A}}
\end{figure}

\noindent One of the big advantages of the ansatz (\ref{eq:ansatz A})
is that as long as $\Phi\left(y\right)$ is non-singular, $\bar{H}\left(y\right)$
is guaranteed to be non-singular. We therefore only need to care about
the singularity of $\Phi\left(y\right)$. Another advantage of the
ansatz (\ref{eq:ansatz A}) is that every individual term inside the
log is non-singular, in contrast to the perturbative solution where
all terms are singular. Singularities appear if and only if
\begin{equation}
\sum_{k=1}^{\infty}\frac{\lambda_{k}}{1+\gamma^{2k}}=0,\label{eq:singular condition A}
\end{equation}
has real roots. In the perturbative regime $\left|y\right|\to\infty$
($\alpha^{\prime}\to0$), the ansatz $\Phi\left(x\right)$ is expanded
as,
\begin{eqnarray}
\Phi(y/\sqrt{\alpha^{\prime}}\to\infty) & = & \frac{1}{2}\log\left(\frac{\lambda_{1}}{\gamma^{2}}\right)+\frac{1}{2}\log\left(\sum_{k=1}^{\infty}\frac{1}{\gamma^{2k-2}}\frac{\lambda_{k}/\lambda_{1}}{1+1/\gamma^{2k}}\right)\nonumber \\
 & = & \frac{1}{2}\log\left(\frac{\lambda_{1}}{\gamma^{2}}\right)+\frac{1}{2}\log\left(\frac{1}{1+1/\gamma^{2}}+\sum_{k=2}^{\infty}\frac{1}{\gamma^{2k-2}}\frac{\lambda_{k}/\lambda_{1}}{1+1/\gamma^{2k}}\right)\nonumber \\
 & = & -\frac{1}{2}\log\left(\frac{\gamma^{2}}{\lambda_{1}}\right)+\frac{\lambda_{2}-\lambda_{1}}{2\lambda_{1}}\frac{1}{\gamma^{2}}+\frac{\lambda_{1}^{2}+2\left(\lambda_{2}+\lambda_{3}\right)\lambda_{1}-\lambda_{2}^{2}}{4\lambda_{1}^{2}}\frac{1}{\gamma^{4}}\nonumber \\
 &  & -\frac{\lambda_{1}^{3}+3\left(\lambda_{2}-\lambda_{3}-\lambda_{4}\right)\lambda_{1}^{2}+3\lambda_{2}\left(\lambda_{2}+\lambda_{3}\right)\lambda_{1}-\lambda_{2}^{3}}{6\lambda_{1}^{3}}\frac{1}{\gamma^{6}}+\cdots.
\end{eqnarray}
To match the perturbative solution (\ref{eq:perturbative solution}),
the coefficients $\lambda_{i}$ are fixed:

\begin{eqnarray}
\lambda_{1} & = & \frac{1}{\beta^{2}},\quad\lambda_{2}=\frac{2}{\beta^{2}},\quad\lambda_{3}=\frac{4+512\bar{c}_{3}}{3\beta^{2}},\quad\lambda_{4}=\frac{8}{15\beta^{2}}\left(3136\bar{c}_{3}-1536\bar{c}_{4}+23\right),\nonumber \\
\lambda_{5} & = & \frac{8\left(1638400\bar{c}_{3}^{2}+66688\bar{c}_{3}-53248\bar{c}_{4}+20480\bar{c}_{5}+219\right)}{35\beta^{2}},\nonumber \\
 & \cdots & ,\label{eq:lambda fixing}
\end{eqnarray}
where we used $\bar{c}_{2}=-1/64$. In order to show the features
of this solution more clearly, we require the solutions (\ref{eq:ansatz A})
and (\ref{eq:general H}) to cover the first two terms of (\ref{eq:perturbative solution})
in the perturbative regime, because the current string theory only
gives the first two coefficients of $\alpha^{\prime}$ corrections,
$\bar{c}_{1}=c_{1}=-\frac{1}{8}$ and $\bar{c}_{2}=-c_{2}=-\frac{1}{64}$.
Although this example is special, it contains all the features of
$\alpha^{\prime}$ corrections. To cover $\bar{c}_{1}$ and $\bar{c}_{2}$,
we only need to keep two coefficients $\lambda_{1}$ and $\lambda_{2}$.
Therefore, the solution (\ref{eq:ansatz A}) becomes

\begin{equation}
\Phi\left(y\right)=\frac{1}{2}\log\left(\frac{1}{4d}\frac{\alpha^{\prime}}{\alpha^{\prime}+2dy^{2}}+\frac{1}{2d}\frac{\alpha^{\prime2}}{\alpha^{\prime2}+4d^{2}y^{4}}\right),\label{eq:simplest solution}
\end{equation}

\noindent where $\beta^{2}=4d$ to match the perturbative solutions.
The $O\left(d,d\right)$ invariant dilaton $\Phi\left(y\right)$ and
its associated Hubble-like parameters $\bar{H}\left(y\right)$ are
presented in Fig. (\ref{fig:solution A})

\subsection*{Solution B}

Supposing the coefficients $c_{k\le n}$ are known, another interesting
ansatz is given by

\begin{equation}
\Phi\left(y\right)=-\frac{1}{2N}\log\left[\stackrel[k=0]{N}{\sum}\rho_{k}\gamma^{2k}\right],\qquad\gamma\equiv\frac{y}{y_{0}}=\frac{\sqrt{2d}}{\sqrt{\alpha^{\prime}}}y,\label{eq:ansatz B}
\end{equation}
where $N\ge n-1$ is some arbitrary integer and $\rho_{0}>0$, $\rho_{N}>0$.
Also from Eq. (\ref{eq:EOM in dilaton}), we have

\noindent
\begin{eqnarray}
\bar{H}\left(y\right) & = & \frac{-\left(\sum\limits _{k=0}^{N}\rho_{k}\gamma^{2k}\right)\sum\limits _{k=0}^{N}2k\left(2k-1\right)\rho_{k}\gamma^{2k-2}+\left(\sum\limits _{k=0}^{N}2k\rho_{k}\gamma^{2k-1}\right){}^{2}}{\sqrt{2}\sqrt{\alpha^{\prime}}\beta N\left(\sum\limits _{k=0}^{N}\rho_{k}\gamma^{2k}\right){}^{2-\frac{1}{2N}}},\nonumber \\
\bar{f}\left(\bar{H}\left(y\right)\right) & = & -\frac{2\sqrt{2}\beta d}{\sqrt{\alpha^{\prime}}}\left(\sum_{k=0}^{N}\rho_{k}\gamma^{2k}\right)^{-\frac{1}{2N}},\nonumber \\
\bar{g}\left(\bar{H}\left(y\right)\right) & = & -\frac{d\left(\sum\limits _{k=0}^{N}2k\rho_{k}\gamma^{2k-1}\right)^{2}}{2\alpha^{\prime}N^{2}\left(\sum\limits _{k=0}^{N}\rho_{k}\gamma^{2k}\right)^{2}}.\label{eq:general H B}
\end{eqnarray}

\noindent In the perturbative regime $\left|y\right|\to\infty$ ($\alpha^{\prime}\to0$),
the ansatz $\Phi\left(y\right)$ in (\ref{eq:ansatz B}) can be expanded
as

\begin{eqnarray}
\Phi\left(y\right) & = & -\frac{1}{2}\log\left(\gamma^{2}\rho_{N}^{1/N}\right)-\frac{1}{2N}\Bigg\{\frac{\rho_{N-1}}{\rho_{N}}\frac{1}{\gamma^{2}}+\frac{2\rho_{N}\rho_{N-2}-\rho_{N-1}^{2}}{2\rho_{N}^{2}}\frac{1}{\gamma^{4}}+\frac{3\rho_{N}^{2}\rho_{N-3}-3\rho_{N}\rho_{N-1}\rho_{N-2}+\rho_{N-1}^{3}}{3\rho_{N}^{3}}\frac{1}{\gamma^{6}}\nonumber \\
 &  & +\frac{4\rho_{N}^{3}\rho_{N-4}-2\rho_{N}^{2}\rho_{N-2}^{2}-4\rho_{N}^{2}\rho_{N-1}\rho_{N-3}+4\rho_{N}\rho_{N-1}^{2}\rho_{N-2}-\rho_{N-1}^{4}}{4\rho_{N}^{4}}\frac{1}{\gamma^{8}}+{\cal O}\left(\frac{1}{\gamma^{10}}\right)\Bigg\}.
\end{eqnarray}
To match the perturbative solution (\ref{eq:perturbative solution}),
the coefficients $\rho_{i}$ are fixed as:
\begin{eqnarray}
\rho_{N} & = & \beta^{2N},\quad\rho_{N-1}=-N\beta^{2N},\quad\rho_{N-2}=\frac{\beta^{2N}}{12}N\left(6N-2048\bar{c}_{3}-22\right),\quad\cdots,
\end{eqnarray}
where we use $\bar{c}_{2}=-1/64$.

It is worth noting that these classes of solutions (\ref{eq:ansatz B})
and (\ref{eq:general H B}) are very special. They are not always
non-singular, and the results depend on the specific values of $\bar{c}_{k}$.
For example, consider the simplest case that solutions (\ref{eq:ansatz B})
and (\ref{eq:general H B}) only cover the first two terms of eq.
(\ref{eq:perturbative solution}) in the perturbative regime $|y|\to\infty$
(or equivalently $\alpha^{\prime}\to0$). In other words, we only
need to keep $\rho_{0}$ and $\rho_{1}$ in (\ref{eq:ansatz B}) and
(\ref{eq:general H B}). The solution is

\begin{equation}
\Phi\left(y\right)=-\frac{1}{2}\log\left[4d\left(-1+\frac{2d}{\alpha^{\prime}}y^{2}\right)\right].
\end{equation}

\noindent It is not difficult to check that $\Phi\left(y\right)$
has a singularity, and therefore the corresponding $\bar{H}\left(y\right)$
is singular:

\begin{equation}
\bar{H}\left(y\right)=\frac{\sqrt{2}\left(2dy^{2}+\alpha^{\prime}\right)}{\left(2dy^{2}-\alpha^{\prime}\right)^{3/2}}.
\end{equation}

\begin{figure}[H]
\noindent \begin{centering}
\includegraphics[scale=0.45]{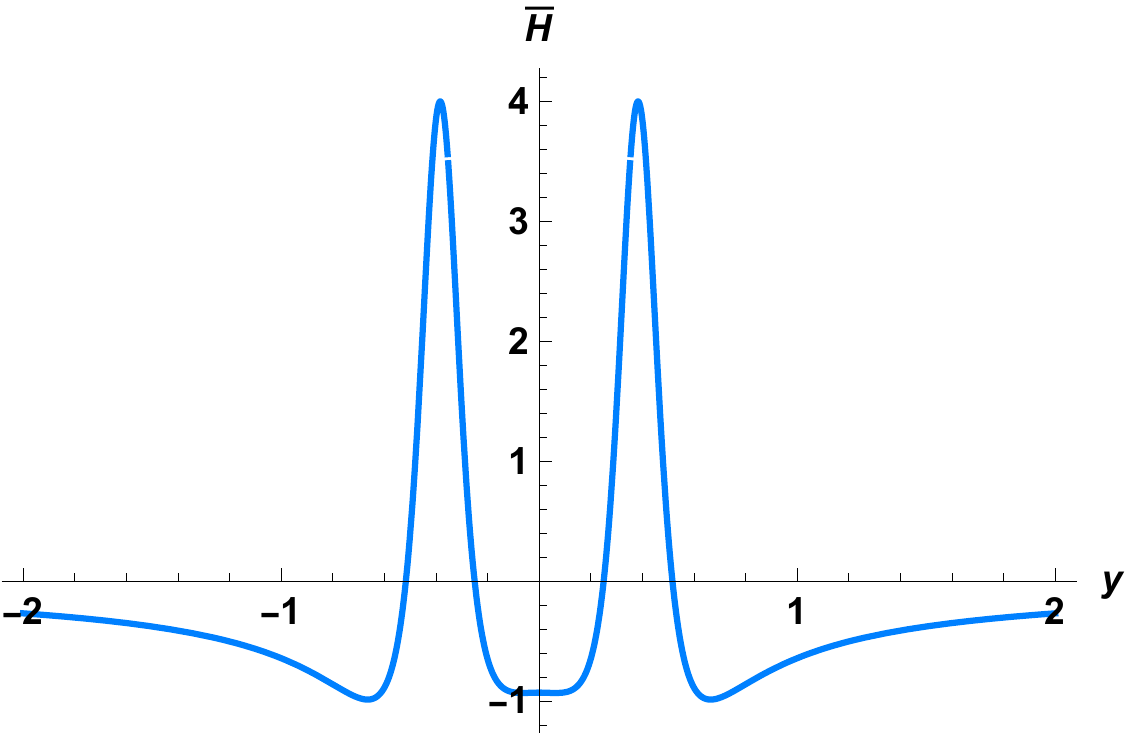}$\qquad$\includegraphics[scale=0.45]{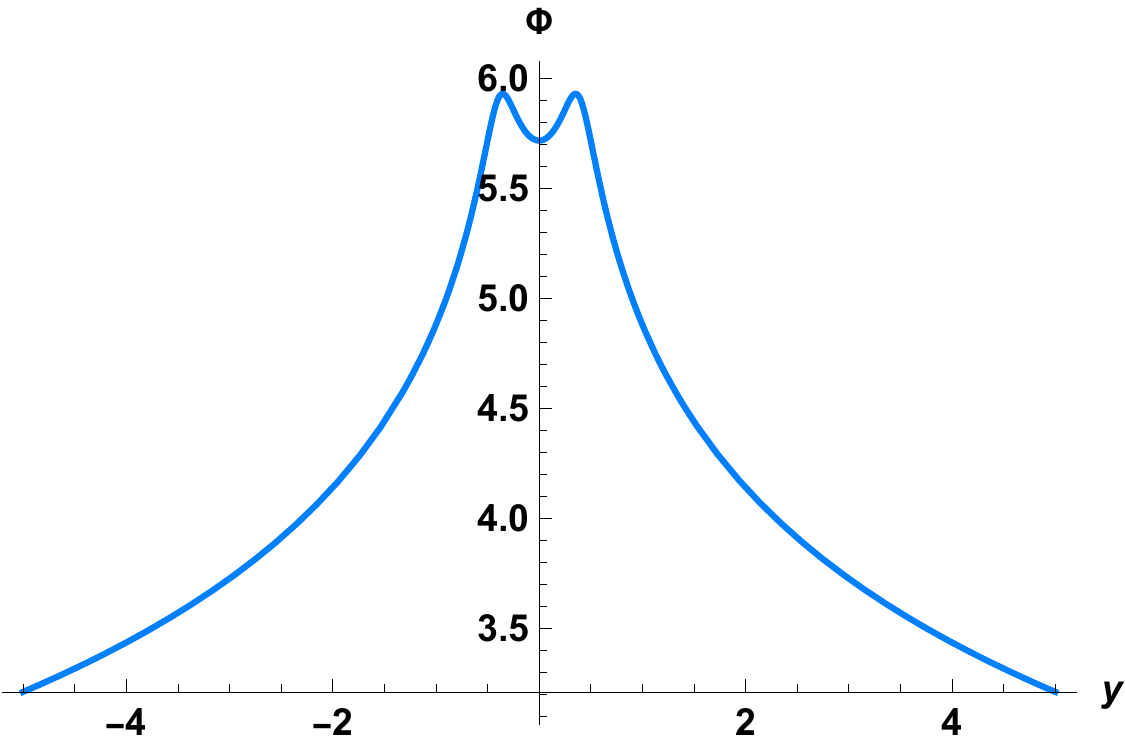}
\par\end{centering}
\caption{The left panel shows the non-perturbative Hubble-like parameters $\bar{H}\left(y\right)$
vs. coordinate of  dimension $y$. The right panel presents the $O\left(d,d\right)$
dilaton $\Phi\left(y\right)$ along the  dimension $y$. We set parameters
$d=4$ and $\alpha^{\prime}=1$.\label{fig:solution B}}
\end{figure}

\noindent To get a non-singular solution, the number of coefficients
$\rho_{i}$ must be more than two, namely we need to know $\bar{c}_{3}$
and $\rho_{3}$ . Considering the simplest case where $N=2$, the
Hubble parameter becomes
\begin{equation}
\bar{H}\left(y\right)=\frac{\sqrt{2}3^{3/4}\left(5\alpha^{\prime3}-1024\alpha^{\prime2}\bar{c}_{3}\left(6dy^{2}-\alpha^{\prime}\right)-24d^{3}y^{6}+12\alpha^{\prime}d^{2}y^{4}-42\alpha^{\prime2}dy^{2}\right)}{\left(-5\alpha^{\prime2}-1024\alpha^{\prime2}\bar{c}_{3}+12d^{2}y^{4}-12\alpha^{\prime}dy^{2}\right){}^{7/4}}.
\end{equation}
To have a non-singular solution for all regions of $y$, there must
be $\bar{c}_{3}<-1/128$. Moreover, in the future, even if $\bar{c}_{3}$
is determined by other theories and the condition is violated, we
can still obtain solutions without singularity by taking $\rho_{4}$
which may put a constraint on $\bar{c}_{4}$, and so on. In the case
of $N=2$, the form of the relevant Ricci scalar is still very complicated.
As an alternative, we give Fig. (\ref{fig:solution B}) to demonstrate
the Hubble-like parameters $\bar{H}\left(y\right)$.

\noindent

\subsection*{Solutions in Einstein frame}

To study the physical properties of the solutions, we may transform
our regular solutions from the string frame to the Einstein frame.
Let us recall the relation between string frame and Einstein frame:

\begin{equation}
g_{\mu\nu}^{E}=\exp\left(-\frac{4\phi}{d-1}\right)g_{\mu\nu}.
\end{equation}

\noindent Therefore, it is straightforward to get the solutions in
the Einstein frame by:

\begin{equation}
\bar{H}_{\pm}^{E}\left(y\right)=\frac{\bar{a}_{E}^{\prime}\left(y\right)}{\bar{a}_{E}\left(y\right)}=-\frac{1}{d-1}\left(\Phi^{\prime}\left(y\right)+\bar{H}_{\pm}\left(y\right)\right).
\end{equation}

\noindent To verify the singularities of these solutions, we need
to consider Kretschmann scalar, which can be given by

\noindent
\begin{equation}
R_{\mu\nu\rho\sigma}R^{\mu\nu\rho\sigma}=4dH_{E}^{\prime2}+8dH_{E}^{\prime}H_{E}^{2}+2\left(d+1\right)dH_{E}^{4},
\end{equation}

\noindent The following pictures show the Kretschmann scalars of $\bar{H}_{+}^{E}$
for solutions A and B:
\begin{figure}[H]
\noindent \begin{centering}
\includegraphics[scale=0.45]{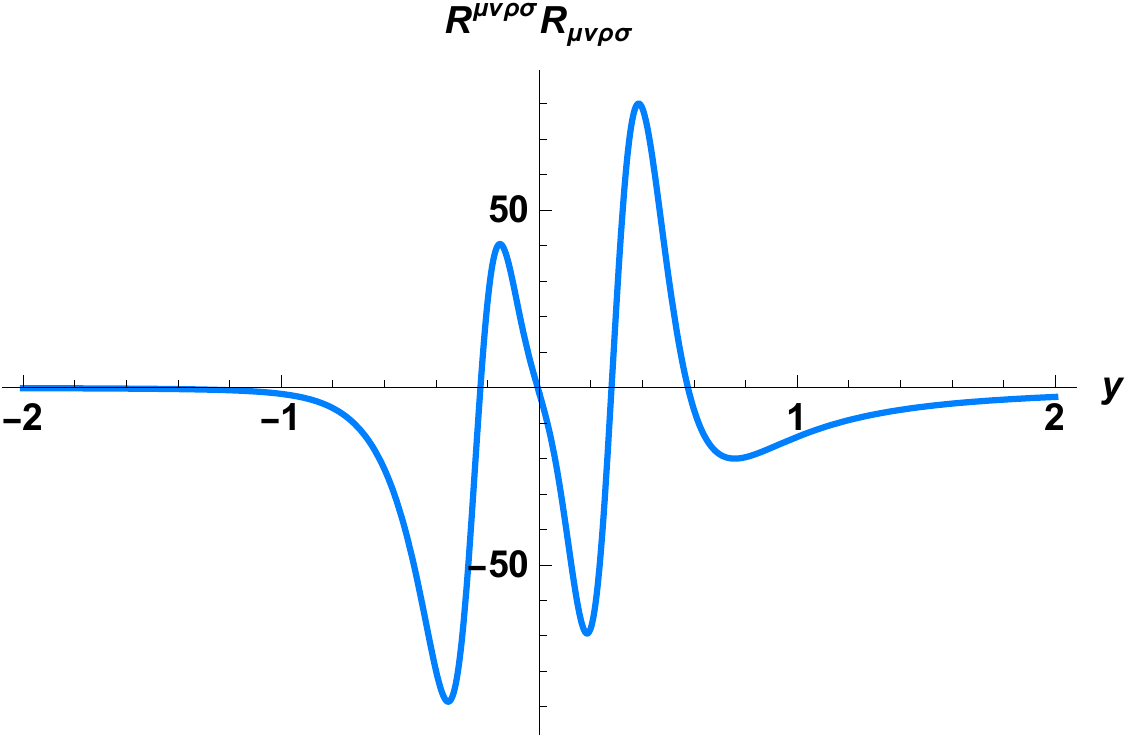}$\qquad$\includegraphics[scale=0.45]{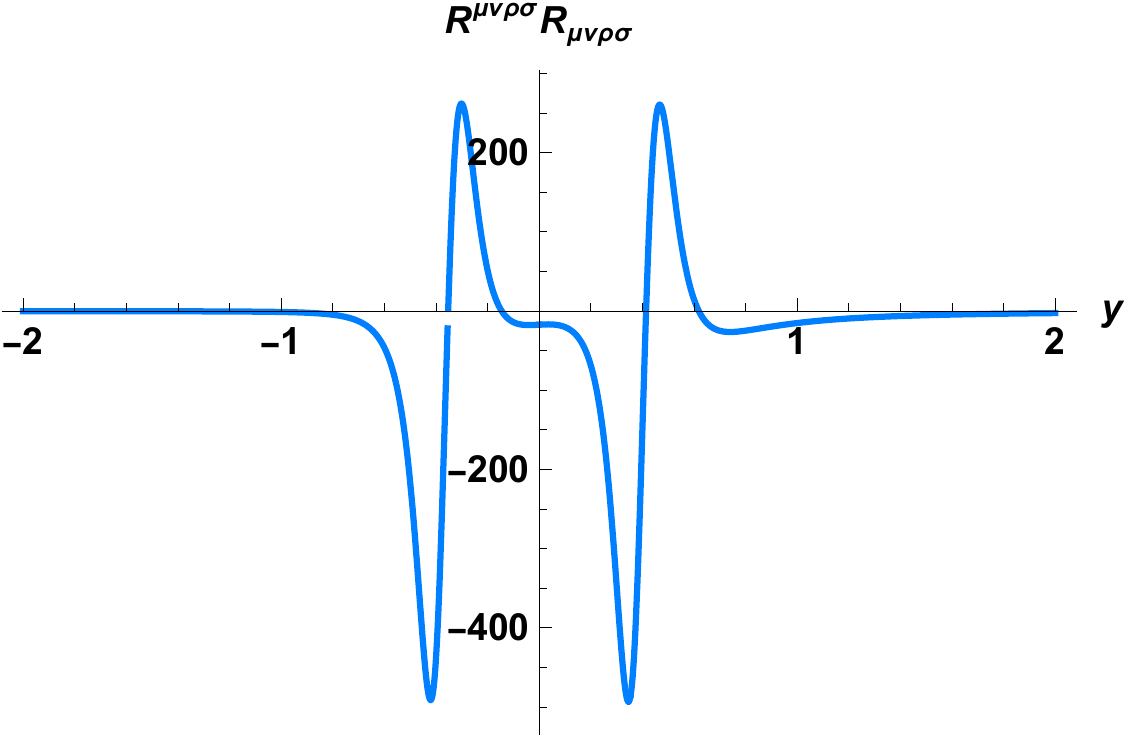}
\par\end{centering}
\caption{The left panel shows the Kretschmann scalar of solution A vs. coordinate
of dimension $y$. The right panel presents the Kretschmann scalar
of solution B vs. coordinate of dimension $y$. We set parameters
$d=4$ and $\alpha^{\prime}=1$.\label{fig:Einstein frame}}
\end{figure}

\section{Concluding remarks}

In this paper, we provided systematic methods to construct two classes
of non-singular non-perturbative solutions, which matched the perturbative
solutions to an arbitrary order in $\alpha^{\prime}$. The methods
depended on the Hohm-Zwiebach action \cite{Hohm:2015doa,Hohm:2019ccp,Hohm:2019jgu}.
The results implied that the complete $\alpha^{\prime}$ corrections
were able to remove naked singularities of spacetime, and then formed
non-singular backgrounds.

In addition, if we consider all $\alpha^{\prime}$ corrections as
an effective potential, say

\begin{eqnarray}
S & = & \int d^{d+1}x\sqrt{-g}e^{-2\phi}\left[R+4\left(\partial_{\mu}\phi\right)^{2}+\frac{\alpha^{\prime}}{4}R_{\mu\nu\rho\sigma}R^{\mu\nu\rho\sigma}+\ldots\right]\nonumber \\
 & = & \int d^{d+1}x\sqrt{-g}e^{-2\phi}\left[R+4\left(\partial_{\mu}\phi\right)^{2}-V\left(\phi\right)\right],\label{eq:effective potential action}
\end{eqnarray}

\noindent it is possible to figure out the effective potential $V\left(\phi\right)$
by comparing actions (\ref{eq:effective potential action}) with (\ref{eq: Odd with alpha x}).
The result is

\begin{equation}
V\left(y\right)=\left(\bar{g}\left(\bar{H}\left(y\right)\right)+d\bar{H}\left(y\right)^{2}\right)-\bar{H}\left(\bar{f}\left(\bar{H}\left(y\right)\right)+2d\bar{H}\left(y\right)\right).
\end{equation}

\begin{figure}[H]
\begin{centering}
\includegraphics[scale=0.35]{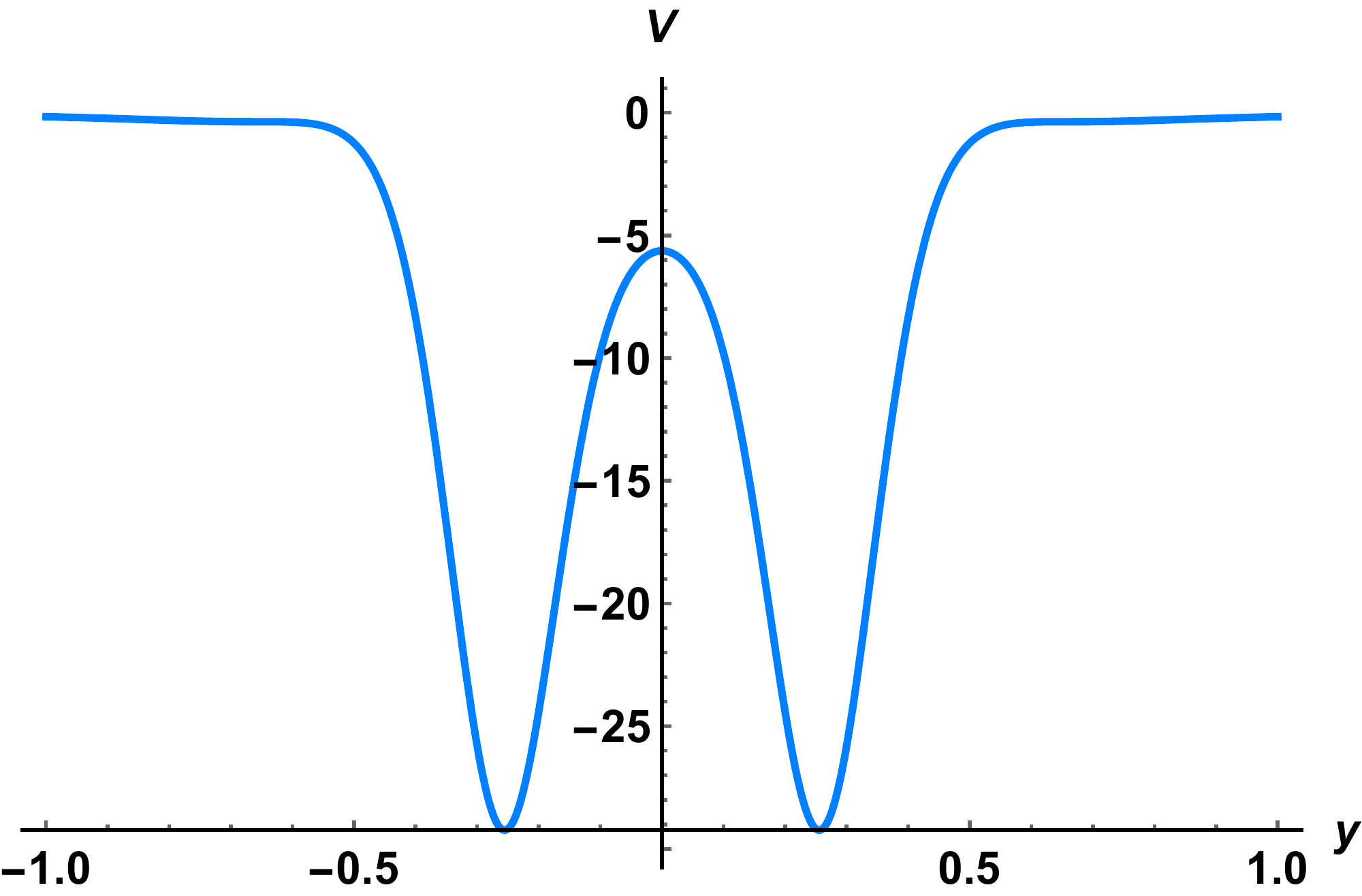}
\par\end{centering}
\centering{}\caption{\label{fig:effective potential A} effective potential $V\left(y\right)$
vs. coordinate of  dimension $y$. We set $d=4$ and $\alpha^{\prime}=1$
in this plot).}
\end{figure}

\noindent For example, the effective potential for solution A (\ref{eq:ansatz A})
can be plotted in Fig. (\ref{fig:effective potential A}).

In the near future, it is worth discussing how to define the junction
functions, tensor perturbations and graviton localization in the Hohm-Zwiebach
action. Then, it is possible to study the non-perturbative stringy
effects in the braneworlds.

\vspace{5mm}

{\bf Acknowledgements}
We thank the useful discussions with Xin Li, Yu-Xiao Liu, Peng Wang, Houwen Wu, Haitang Yang, Hao Yu and Yuan Zhong. This work is supported in part by the NSFC (Grant No. 12105031).


\begin{thebibliography}{99}

\bibitem{Tseytlin:1991wr}    A.~A.~Tseytlin,   ``Duality and dilaton,''   Mod.\ Phys.\ Lett.\ A {\bf 6}, 1721 (1991).   doi:10.1142/S021773239100186X   

\bibitem{Veneziano:1991ek}
G.~Veneziano,   ``Scale factor duality for classical and quantum strings,''   Phys.\ Lett.\ B {\bf 265}, 287 (1991).   doi:10.1016/0370-2693(91)90055-U   

\bibitem{Meissner:1991zj}   K.~A.~Meissner and G.~Veneziano,   ``Symmetries of cosmological superstring vacua,''   Phys.\ Lett.\ B {\bf 267}, 33 (1991).   doi:10.1016/0370-2693(91)90520-Z   

\bibitem{Sen:1991zi}    A.~Sen,   ``O(d) x O(d) symmetry of the space of cosmological solutions in string theory, scale factor duality and two-dimensional black holes,''   Phys.\ Lett.\ B {\bf 271}, 295 (1991).   doi:10.1016/0370-2693(91)90090-D   

\bibitem{Sen:1991cn}   A.~Sen,   ``Twisted black p-brane solutions in string theory,''   Phys.\ Lett.\ B {\bf 274}, 34 (1992)   doi:10.1016/0370-2693(92)90300-S   [hep-th/9108011].   

\bibitem{Tseytlin:1991xk}    A.~A.~Tseytlin and C.~Vafa,   ``Elements of string cosmology,''   Nucl.\ Phys.\ B {\bf 372}, 443 (1992)   doi:10.1016/0550-3213(92)90327-8   [hep-th/9109048].   





\bibitem{Veneziano:2000pz}    G.~Veneziano,   ``String cosmology: The Pre - big bang scenario,''   doi:10.1007/3-540-45334-2-12   [hep-th/0002094].   


\bibitem{Gasperini:2002bn}    M.~Gasperini and G.~Veneziano,   ``The Pre - big bang scenario in string cosmology,''   Phys.\ Rept.\  {\bf 373}, 1 (2003)   doi:10.1016/S0370-1573(02)00389-7   [hep-th/0207130].   

\bibitem{Gasperini:2007vw}    M.~Gasperini and G.~Veneziano,   ``String Theory and Pre-big bang Cosmology,''   Nuovo Cim.\ C {\bf 38}, no. 5, 160 (2016)   doi:10.1393/ncc/i2015-15160-8   [hep-th/0703055].   

\bibitem{Gasperini:1992em}    M.~Gasperini and G.~Veneziano,   ``Pre - big bang in string cosmology,''   Astropart.\ Phys.\  {\bf 1}, 317 (1993)   doi:10.1016/0927-6505(93)90017-8   [hep-th/9211021].   




\bibitem{Gasperini:2003pb}    M.~Gasperini, M.~Giovannini and G.~Veneziano,   ``Perturbations in a nonsingular bouncing universe,''   Phys.\ Lett.\ B {\bf 569}, 113 (2003)   doi:10.1016/j.physletb.2003.07.028   [hep-th/0306113].   



\bibitem{Gasperini:2004ss}   M.~Gasperini, M.~Giovannini and G.~Veneziano,   ``Cosmological perturbations across a curvature bounce,''   Nucl.\ Phys.\ B {\bf 694} (2004) 206   doi:10.1016/j.nuclphysb.2004.06.020   [hep-th/0401112].   

\bibitem{Gasperini:book} Maurizio Gasperini. Elements of String Cosmology. Cambridge University Press, 2007.








\bibitem{Meissner:1996sa}    K.~A.~Meissner,   ``Symmetries of higher order string gravity actions,''   Phys.\ Lett.\ B {\bf 392}, 298 (1997)   doi:10.1016/S0370-2693(96)01556-0   [hep-th/9610131].   




\bibitem{Hohm:2015doa}    O.~Hohm and B.~Zwiebach,   ``T-duality Constraints on Higher Derivatives Revisited,''   JHEP {\bf 1604}, 101 (2016)   doi:10.1007/JHEP04(2016)101   [arXiv:1510.00005 [hep-th]].   

\bibitem{Hohm:2019ccp}    O.~Hohm and B.~Zwiebach,   ``Non-perturbative de Sitter vacua via $\alpha'$ corrections,''   arXiv:1905.06583 [hep-th].   

\bibitem{Hohm:2019jgu}    O.~Hohm and B.~Zwiebach,   ``Duality Invariant Cosmology to all Orders in $\alpha'$,''   arXiv:1905.06963 [hep-th].   


\bibitem{Wang:2019mwi}    P.~Wang, H.~Wu and H.~Yang,   ``Are nonperturbative AdS vacua possible in sonic string theory?,''   Phys.\ Rev.\ D {\bf 100}, no. 4, 046016 (2019)   doi:10.1103/PhysRevD.100.046016   [arXiv:1906.09650 [hep-th]].   


\bibitem{Penrose:1969pc} R.~Penrose, ``Gravitational collapse: The role of general relativity,'' Riv. Nuovo Cim. \textbf{1}, 252-276 (1969) doi:10.1023/A:1016578408204 



\bibitem{Kar:1998rv} S.~Kar, ``Naked singularities in low-energy, effective string theory,'' Class. Quant. Grav. \textbf{16}, 101-115 (1999) doi:10.1088/0264-9381/16/1/008 [arXiv:hep-th/9804039 [hep-th]]. 



\bibitem{Wang:2019kez}    P.~Wang, H.~Wu, H.~Yang and S.~Ying,   ``Non-singular string cosmology via $\alpha^{\prime}$ corrections,''   JHEP {\bf 1910}, 263 (2019)   doi:10.1007/JHEP10(2019)263   [arXiv:1909.00830 [hep-th]].   

\bibitem{Wang:2019dcj}    P.~Wang, H.~Wu, H.~Yang and S.~Ying,   ``Construct $\alpha^{\prime}$ corrected or loop corrected solutions without curvature singularities,''   JHEP {\bf 2001}, 164 (2020)   doi:10.1007/JHEP01(2020)164   [arXiv:1910.05808 [hep-th]].   


\bibitem{Wang:2020eln} P.~Wang, H.~Wu, H.~Yang and S.~Ying, ``Derive Lovelock Gravity from String Theory in Cosmological Background,'' JHEP \textbf{05}, 218 (2021) doi:10.1007/JHEP05(2021)218 [arXiv:2012.13312 [hep-th]]. 

\bibitem{Bernardo:2019bkz} H.~Bernardo, R.~Brandenberger and G.~Franzmann, ``O$(d,d)$ covariant string cosmology to all orders in $\alpha^{\prime}$,'' JHEP \textbf{02}, 178 (2020) doi:10.1007/JHEP02(2020)178 [arXiv:1911.00088 [hep-th]]. 


\bibitem{Wu:2013sha}
H.~Wu and H.~Yang, ``Double Field Theory Inspired Cosmology,'' JCAP \textbf{07}, 024 (2014) doi:10.1088/1475-7516/2014/07/024 [arXiv:1307.0159 [hep-th]]. 


\bibitem{Lescano:2021nju} E.~Lescano and N.~Mir\'on-Granese, ``Double Field Theory with matter and its cosmological application,'' [arXiv:2111.03682 [hep-th]]. 


\bibitem{Dzhunushaliev:2009va} V.~Dzhunushaliev, V.~Folomeev and M.~Minamitsuji, ``Thick brane solutions,'' Rept. Prog. Phys. \textbf{73}, 066901 (2010) doi:10.1088/0034-4885/73/6/066901 [arXiv:0904.1775 [gr-qc]]. 

\bibitem{Liu:2017gcn} Y.~X.~Liu, ``Introduction to Extra Dimensions and Thick Braneworlds,'' doi:10.1142/9789813237278\_0008 [arXiv:1707.08541 [hep-th]]. 

\bibitem{Zwiebach:1985uq}
B.~Zwiebach, ``Curvature Squared Terms and String Theories,'' Phys. Lett. B \textbf{156}, 315-317 (1985) doi:10.1016/0370-2693(85)91616-8 








































\end{thebibliography}
\end{document}